\documentclass[10pt,conference]{IEEEtran}
\IEEEoverridecommandlockouts

\usepackage{cite}
\usepackage{amsmath,amssymb,amsfonts}
\usepackage{algorithm}
\usepackage[noend]{algpseudocode}
\usepackage{etoolbox}
\usepackage{graphicx}
\usepackage{textcomp}
\usepackage{xcolor}

\usepackage{array}
\usepackage[caption=false,font=footnotesize,labelfont=sf,textfont=sf]{subfig}
\usepackage{textcomp}
\usepackage{stfloats}
\usepackage{url}
\usepackage{verbatim}
\usepackage{lipsum}
\usepackage{graphicx}
\usepackage{epstopdf}

\usepackage{cite}

\def\BibTeX{{\rm B\kern-.05em{\sc i\kern-.025em b}\kern-.08em
    T\kern-.1667em\lower.7ex\hbox{E}\kern-.125emX}}

\makeatletter

\def\therule{\makebox[\algorithmicindent][l]{\hspace*{.5em}\vrule height .75\baselineskip depth .25\baselineskip}}%

\newtoks\therules
\therules={}
\def\appendto#1#2{\expandafter#1\expandafter{\the#1#2}}
\def\gobblefirst#1{
  #1\expandafter\expandafter\expandafter{\expandafter\@gobble\the#1}}%
\def\LState{\State\unskip\the\therules}
\def\pushindent{\appendto\therules\therule}%
\def\popindent{\gobblefirst\therules}%
\def\printindent{\unskip\the\therules}%
\def\printandpush{\printindent\pushindent}%
\def\popandprint{\popindent\printindent}%

\algdef{SE}[WHILE]{While}{EndWhile}[1]
  {\printandpush\algorithmicwhile\ #1\ \algorithmicdo}
  {\popandprint\algorithmicend\ \algorithmicwhile}%
\algdef{SE}[FOR]{For}{EndFor}[1]
  {\printandpush\algorithmicfor\ #1\ \algorithmicdo}
  {\popandprint\algorithmicend\ \algorithmicfor}%
\algdef{S}[FOR]{ForAll}[1]
  {\printindent\algorithmicforall\ #1\ \algorithmicdo}%
\algdef{SE}[LOOP]{Loop}{EndLoop}
  {\printandpush\algorithmicloop}
  {\popandprint\algorithmicend\ \algorithmicloop}%
\algdef{SE}[REPEAT]{Repeat}{Until}
  {\printandpush\algorithmicrepeat}[1]
  {\popandprint\algorithmicuntil\ #1}%
\algdef{SE}[IF]{If}{EndIf}[1]
  {\printandpush\algorithmicif\ #1\ \algorithmicthen}
  {\popandprint\algorithmicend\ \algorithmicif}%
\algdef{C}[IF]{IF}{ElsIf}[1]
  {\popandprint\pushindent\algorithmicelse\ \algorithmicif\ #1\ \algorithmicthen}%
\algdef{Ce}[ELSE]{IF}{Else}{EndIf}
  {\popandprint\pushindent\algorithmicelse}%
\algdef{SE}[PROCEDURE]{Procedure}{EndProcedure}[2]
   {\printandpush\algorithmicprocedure\ \textproc{#1}\ifthenelse{\equal{#2}{}}{}{(#2)}}%
   {\popandprint\algorithmicend\ \algorithmicprocedure}%
\algdef{SE}[FUNCTION]{Function}{EndFunction}[2]
   {\printandpush\algorithmicfunction\ \textproc{#1}\ifthenelse{\equal{#2}{}}{}{(#2)}}%
   {\popandprint\algorithmicend\ \algorithmicfunction}%
\makeatother

\begin{document}
\begin{sloppypar}
\title{Intelligent Multi-link EDCA Optimization for Delay-Bounded QoS in Wi-Fi 7\\
}
\author{
    \IEEEauthorblockN{Peini Yi$^*$, Wenchi Cheng$^*$, Jingqing Wang$^*$, Jinzhe Pan$^{\dag}$, Yuehui Ouyang$^{\dag}$ and Wei Zhang$^{\ddag}$}
    \IEEEauthorblockA{$^*$\textit{School of Telecommunication Engineering, Xidian University, Xi'an, China }\\
    $^{\dag}$\textit{Wireless Communication Department, Honor Device Co. Ltd., Shenzhen, China}\\
    $^{\ddag}$\textit{School of Electrical Engineering and Telecommunications, University of New South Wales, Sydney, NSW}\\
    Email: $^*$pnyi@stu.xidian.edu.cn,$^*$\{jqwangxd, wccheng\}@xidian.edu.cn,\\
    $^{\dag}$\{panjinzhe, yuehuiouyang\}@honor.com, $^{\ddag}$w.zhang@unsw.edu.au.
    }
    \thanks{This work was supported by the National Key Research and Development Program of China under Grant 2024YFC3016000. }
}

\maketitle

\begin{abstract}
    IEEE 802.11be (Wi-Fi 7) introduces Multi-Link Operation (MLO) as a  While MLO offers significant parallelism and capacity, realizing its full potential in guaranteeing strict delay bounds and optimizing Quality of Service (QoS) for diverse, heterogeneous traffic streams in complex multi-link scenarios remain a significant challenge. This is largely due to the limitations of static Enhanced Distributed Channel Access (EDCA) parameters and the complexity inherent in cross-link traffic management. To address this, this paper investigates the correlation between overall MLO QoS indicators and the configuration of EDCA parameters and Acess Catagory (AC) traffic allocation among links. Based on this analysis, we formulate a constrained optimization problem aiming to minimize the sum of overall packet loss rates for all access categories while satisfying their respective overall delay violation probability constraints. A Genetic Algorithm (GA)-based MLO EDCA QoS optimization algorithm is designed to efficiently search the complex configuration space of AC assignments and EDCA parameters. Experimental results demonstrate that the proposed approach's efficacy in generating adaptive MLO configuration strategies that align with diverse service requirements. The proposed solution significantly improves delay distribution characteristics, and enhance QoS robustness and resource utilization efficiency in high-load MLO environments. 
\end{abstract}

\begin{IEEEkeywords}
MLO, EDCA, QoS, delay bound, genetic algorithm.
\end{IEEEkeywords}

\section{Introduction}
\IEEEPARstart{T}{he} IEEE 802.11be standard (Wi-Fi 7) represents a transformative advancement in wireless local area networks, driven by the growing demand for high-throughput, low-latency applications such as real-time gaming, augmented reality, and ultra-high-definition video conferencing~\cite{Deng2020}. A cornerstone feature of this standard is Multi-Link Operation (MLO)\cite{lopez-raventosMultiLinkOperationIEEE2022}, which enables Multi-Link Devices (MLDs) to concurrently transmit and receive data across multiple links in distinct frequency bands. This parallelism significantly enhances throughput and supports delay-sensitive traffic. Within the Wi-Fi framework, the Enhanced Distributed Channel Access (EDCA) mechanism ensures Quality of Service (QoS) by providing differentiated access through parameters such as Arbitration Inter-Frame Space (AIFS) and Transmission Opportunity limit (TXOP) for various Access Categories (ACs). While the synergy of MLO’s concurrent capabilities and EDCA’s prioritization offers substantial potential for robust QoS and efficient resource utilization under high load, achieving stringent delay guarantees, particularly for end-to-end delay bounds~\cite{Wang2024, Zhang2022}, remains a critical challenge in complex MLO scenarios.

Ensuring strict end-to-end delay bounds for diverse traffic streams is essential for emerging applications. However, integrating MLO with EDCA~\cite{huang4DMarkovChain2025} poses key challenges to predictable delay performance. First, traffic management across links demands effective load balancing and cross-link interference mitigation. Second, EDCA’s static parameters for its four ACs fail to adapt to dynamic patterns from varying link assignments. Under high loads, this mismatch exacerbates contention and congestion, raising delay violations for sensitive services and risking starvation of low-priority traffic. Consequently, long-tail delay distributions emerge, degrading QoS.

To address these challenges, this paper proposes a novel optimization framework for MLO networks operating in Simultaneous Transmit and Receive (STR) mode. Inspired by advancements in dynamic MAC protocols, such as the AI-MAC framework~\cite{panMACRevivoArtificial2024}, our approach jointly optimizes AC-to-link traffic allocation and per-AC, per-link EDCA parameter configurations. To quantify MLO QoS, we develop an analytical model that precisely calculates delay violation probabilities. We then formulate a constrained optimization problem to minimize packet loss under strict delay constraints. Since this problem is a complex mixed-integer non-linear optimization, we use genetic algorithm (GA) to efficiently find a near-optimal solution for delay-bounded QoS. Our framework ensures robust performance in next-generation Wi-Fi networks.

The rest of this paper is organized as follows. Section II introduces the multi-link EDCA contention model. Section III analyzes the total delay of the multi-link EDCA model. Section IV analyzes the parameter sensitivity of our models and presents QoS optimization and the corresponding algorithm. Section V provides the numerical results. Finally, we conclude this paper in Section VI.

\section{System Model}
We model an IEEE 802.11be MLO Basic Service Set (BSS) with $M$ independent links in distinct frequency bands. The system includes one MLD Access Point (AP) and multiple MLD stations (STAs) with saturated traffic. We consider the STR mode, enabling MLDs to transmit on one link while receiving on another~\cite{lopez-raventosIEEE80211beMultiLink2021a}. Each link uses independent channel access with RTS/CTS to reduce collisions.

\begin{figure}[t]
    \centering
    \includegraphics[width=3.5in]{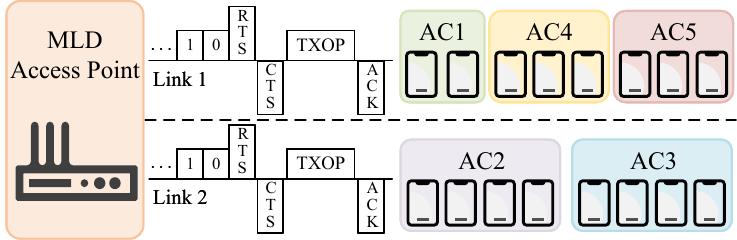}
    \caption{Multi-link EDCA system with $M=2$ links and AC-to-link allocation.}
    \label{fig:MLOEDCA}
\end{figure}

The system supports $I_{\text{all}}$ ACs, where AC$_i$ ($i=1, \ldots, I_{\text{all}}$) is associated with $n_i^{\text{all}}$ STAs, each potentially carrying an uplink and a downlink AC$_i$ flows. We analyze an AC-to-link assignment strategy, mapping all STAs with AC$_i$ flows to a single link $m_i \in \{1, \ldots, M\}$. Fig.~\ref{fig:MLOEDCA} illustrates the multi-link EDCA model for $M=2$, $I_{\text{all}}=5$. 
Each link employs the EDCA mechanism. Single $I (I \leq I_{\text{all}})$ active ACs have STAs assigned, with $n_i = n_i^{\text{all}}$ for AC$_i$ on the link, or $n_i = 0$ otherwise. The EDCA parameters for AC$_i$ are:
\begin{itemize}
    \item CW${_{\min},i}$ and CW${_{\max},i}$: contention window bounds;

    \item AIFS$_i$: inter-frame spacing for priority differentiation;

    \item TXOP$_i$: channel occupancy limit;

    \item R$_i$: retry limit.
\end{itemize}
These ensure prioritized, efficient channel access. We also adjust the retry limit (R$_i$) to improve AC$_i$ frame reliability.

\begin{figure}[t]
    \centering
    \includegraphics[width=2.9in]{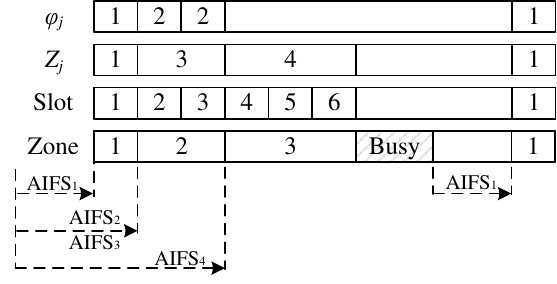}
    \caption{The EDCA contention process within the AIFS zone model on a single link.}
    \label{fig:AIFSZONE}
\end{figure}

To analyze EDCA contention with varying AIFS values among the $I$ active ACs on this single link, we adopt the AIFS zone model \cite{ramaiyanFixedPointAnalysis2008,gaoIEEE80211eEDCA2014}, as illustrated in Fig.~\ref{fig:AIFSZONE}. This model is applied to the $I$ active ACs present on the link. It divides the backoff period following the shortest AIFS among these active ACs (denoted AIFS$_{1}$) into $J$ time zones based on their distinct AIFS values. Each zone $j$ represents a set of time slots where specific active ACs can compete, with $Z_j$ denoting the number of active ACs eligible to contend in zone $j$. For each active AC$_i$ on this link, we define $Z_i^0$ as the first zone where it can begin contending.

The zone for slot $j$ in the backoff period after AIFS$_{1}$ is defined as:
\begin{equation}
\varphi_j = \max\{k \mid h_k < j - 1\},
\end{equation}
where $j = 1, 2, \ldots, h_I$, and $h_k$ is the relative slot offset of the $k$-th AIFS value in the sorted list of the $I$ active ACs' AIFS values on this link, given by:
\begin{equation}
h_k = \frac{\text{AIFS}_k - \text{AIFS}_{1}}{\sigma},
\end{equation}
with $\sigma$ representing the slot time. For the zone model analysis, we assume the AIFS values of the $I$ active ACs on the link are sorted in ascending order (AIFS$_{\min} = \text{AIFS}_1 \leq \cdots \leq \text{AIFS}_{I}$ in the sorted list), and the index $k$ in these equations refers to the index in this sorted list of AIFS values (from $1$ to $I$). Note that the set of active ACs and their corresponding parameters ($n_i$, AIFS$_i$, etc.) are specific to the link being analyzed.

\section{Analytical Model for Multi-Link EDCA QoS}

\subsection{EDCA Collision Model}
For each link, based on the AIFS zone model, we derive fixed-point equations to approximate the collision probability $c_i$ and transmission probability $p_i$ for each $\mathrm{AC}_i$. These probabilities characterize contention dynamics across zones and enable performance evaluation of the EDCA mechanism. 

The stationary probability of the system being in zone $k$, denoted by $\pi_k$, can be computed from the steady-state probabilities of the Markov chain as follows:
\begin{equation}
    \pi_k =
    \begin{cases}
    \pi_0^0 \left[1 + \sum\limits_{j=h_k+2}^{h_{k+1}} q_k^{j - h_k - 1}\right] \alpha_{k-1}, & k \neq J, \\
    \pi_0^0 (1 - q_J)^{-1} \alpha_{J-1}, & k = J,
    \end{cases}
    \label{eq: pi}
\end{equation}
where $\pi_0^0$ is the normalization constant, $\alpha_j$ accounts for transitions between zones, and $q_k$ represents the probability that no station in zone $k$ transmits. These are defined as

\begin{align*}
    \pi_0^0 &= \left(1 + \sum_{i=1}^{J-1} \sum_{j=h_i+2}^{h_{i+1}+1} q_i^{j-h_i-1} \alpha_{i-1} + \frac{q_J \alpha_{J-1}}{1-q_J}\right)^{-1}, \\
    \alpha_j &= \prod_{\ell=1}^{j} q_\ell^{h_{\ell+1} - h_\ell},\quad 
    \alpha_0 = 1, \quad q_k = \prod_{i=1}^{Z_k} r_i^{n_i},
\end{align*}
where \( r_k=1-p_k\) is the probability that AC\(_k\) does not transmit, and $Z_k$ is the number of ACs contending in zone $k$.

Next, we compute the collision probability $c_k$ for AC$_k$, which accounts for collisions in zones where AC$_k$ is eligible to contend (from zone $Z_k^0$ onward):
\begin{equation}
    c_k = \sum_{j=Z^0_k}^{J} \frac{\pi_j}{\sum_{i = Z^0_k}^{J} \pi_i} \left(1 - \frac{\prod_{l=1}^{Z_j}r_\ell^{n_\ell }}{r_k}\right),
    \label{eq: c}
\end{equation}
which weights the collision likelihood in each zone by the normalized probability of AC$_k$ being active.

The transmission probability $p_k$ is derived using a mean-value approximation \cite{xuAccessDelayModel2009}:

\begin{equation}
    p_k=\frac{2}{\eta_k \sum_{j=0}^{R_k-1}c_k^j\left(f_{k,j}-1\right) },
    \label{eq: p}
\end{equation}
where $\eta_k = (1 - c_k)(1 - c_k^{R_k})^{-1}$ reflects the success probability over $R_k$ retransmission attempts, and $f_{k,j} = 2^{\min\{j, m_k\}} \text{CW}_{\min,k}$ defines the contention window size at backoff stage $j$. Here, $m_k = \log_2(\text{CW}_{\max,k} / \text{CW}_{\min,k})$ is the maximum backoff stage, and the factor 2 models the exponential window growth.

Eq.~(\ref{eq: pi})-(\ref{eq: p}) forms a fixed-point system, solved iteratively to obtain $p_k$ and $c_k$ for each AC$_k$. These probabilities enable the calculation of the packet loss probability, given by:
\begin{equation}
    P_{loss,k}  = {c_k}^{R_k},
    \label{eq: Ploss}
\end{equation} 
where packets exceeding the maximum retransmission limit $R_k$ are discarded. This metric underpins the QoS evaluation in subsequent analyses.

\subsection{Delay Performance}
To evaluate delay performance, we model the total delay for AC$_k$. We derive the fundamental timing parameters for the EDCA RTS/CTS access mechanism. These parameters include the data packet transmission time $T_{\text{DATA},k}$ for AC$_k$, and the control frame transmission times $T{_\text{RTS}}$, $T_{\text{CTS}}$, and $T_{\text{ACK}}$, defined as:
\begin{equation}
  T_{\text{DATA},k} = T_{\text{PHY\_H}} + ({L_{\text{MAC\_H}} + L_k})/{r_{\text{data}}}, \
\end{equation}
where $T_{\text{PHY\_H}}$ is the physical layer header duration, $L_{\text{MAC\_H}}$ and $L_k$ denote the MAC header and payload lengths for AC$_k$, respectively, and $L{_\text{RTS}}$, $L_{\text{CTS}}$, and $L_{\text{ACK}}$ represent the lengths of the respective control frames. The data rate $r_{\text{data}}$ and control rate $r_{\text{ctrl}}$ determine the transmission rates for data and control packets, respectively.

A key parameter for delay modeling is the TXOP limit, \( \text{TXOP}_k \). Once \( \text{AC}_k \) obtains channel access, the maximum number of data packets it can transmit within a TXOP is:
\begin{equation}
    N_k = \max\left\{\left\lfloor \frac{\text{TXOP}_k}{\Delta_k} \right\rfloor, 1\right\},
\end{equation}
where $\Delta_k = T_{\text{DATA},k} + T_{\text{ACK}} + 2\text{SIFS}$, and $T_{\text{SIFS}}$ is the Short Interframe Space duration. The floor function ensures at least one packet is sent, even for a short TXOP, accommodating bursty traffic efficiently.

Building on these, we define the collision time $T^{\text{C}}_{k}$ and the successful transmission time $T^{\text{S}}_{k}$ for AC$_k$:
\begin{equation}
    \left\{
    \begin{aligned}
        &T^{\text{C}}_{k} = T_{\text{RTS}} + \text{AIFS}_k,  \\
        &T^{\text{S}}_{k} = T_{\text{RTS}}+ T_{\text{CTS}} + N_{k} \Delta_k + \text{SIFS}+ \text{AIFS}_k,\\
    \end{aligned}
    \right.
\end{equation}
where $T^{\text{C}}_{k}$ reflects the time lost to an RTS collision, and $T^{\text{S}}_{k}$ accounts for the full sequence of RTS, CTS, and $N_k$ DATA-ACK exchanges.

Our objective is to compute the delay violation probability, defined as the probability that the delay of $\mathrm{AC}_k$ exceeds a specified threshold, i.e., $\Pr(D_k \geq D_{\max,k})$, which corresponds to the complementary cumulative distribution function (CCDF) of delay. We adopt a generating function based method from \cite{xuAccessDelayModel2009} to characterize the CCDF of the total delay. The delay generating function for $\mathrm{AC}_k$, denoted by $\widehat{D}(z)$ (with the superscript $(k)$ omitted for clarity), is given as follows:
\begin{equation}
    \label{eq: total_delay}
    \widehat{D}(z) = \frac{1}{N_k} \widehat{A}(z) \widehat{T}(z) \widehat{\epsilon}(z) + \frac{N_k - 1}{N_k} z^{\Delta_{k}/\delta},
\end{equation}
where $z$ is a complex variable, $\delta$ is the discrete time slot duration, and $N_k$ is the number of packets transmitted in a TXOP. The functions $\widehat{A}(z)$, $\widehat{T}(z)$, and $\widehat{\epsilon}(z)$ represent the generating functions for backoff and collision time, successful transmission time, and AIFS defer period, respectively. The second term captures the additional delay for subsequent packets within a TXOP, averaging the data transmission overheads.

The AIFS defer period generating function $\widehat{\epsilon}(z)$ accounts for the AC$_k$ waiting time before contention:
\begin{equation}
\widehat{\epsilon}(z) = 
\begin{cases}
    z^{\mathrm{AIFS}_k / \delta}, & k = 1, \\
    \dfrac{s_{h_k} z^{\mathrm{AIFS}_k / \delta}}{1 - \sum_{\ell=1}^{h_k} \mu_\ell z^{\tau_\ell / \delta}}, & k \neq 1,
\end{cases}
\end{equation}
where $s_{h_k}$ is the probability of the channel remaining idle for $h_k$ slots, $\mu_\ell$ is the probability of an interruption at slot $\ell$, and $\tau_\ell$ is the excess delay due to interruptions, defined as:
\begin{align*}
\tau_{\ell} &= \text{AIFS}_1 +(\ell-1)\sigma + \xi_{\ell} ,\quad s_{\ell} = \prod_{m=1}^{\ell} \prod_{j=1}^{\varphi_m} r_j^{n_j}, \\
\mu_\ell &= \left( \prod_{m=1}^{\ell-1} \prod_{j=1}^{\varphi_m} r_j^{n_j} \right) \left( 1 - \prod_{j=1}^{\varphi_\ell} r_j^{n_j} \right),
\end{align*}

The interruption duration $\xi_\ell$ is:
\begin{equation}
    \begin{aligned}
    &{\xi}_\ell = 
    \begin{cases}
       N_k\Delta_{k}-T_\text{SIFS} , &\text{w.p. } \rho_{k}(\ell),\quad 1\leq k \leq\varphi(\ell), \\
       T_{\text{C}} - \text{AIFS}_k, &\text{w.p. } 1 - \sum_{k=1}^{\varphi(\ell)} \rho_{k}(\ell), \\
    \end{cases} \\
    &\rho_{k}(\ell)=\frac{n_k p_k r_k^{n_k-1}\prod_{j=1,j\neq k}^{\varphi(\ell)}r_j^{n_j}}{1-\prod_{j=1}^{\varphi(\ell)}r_j^{n_j}},
    \end{aligned}
    \end{equation}
where $\rho_k(\ell) $ is the probability that AC$_k$ causes the interruption in zone $\ell$.

The generating function for the backoff and collision time involves the AC$_k$ is:
\begin{equation}
    \widehat{A}(z) = \sum_{i=0}^{R_k - 1} \eta_k c_k^i \widehat{C}(z)^i \prod_{j=0}^i \frac{1 - \widehat{Y}(z)^{f_{k,j}}}{f_{k,j} (1 - \widehat{Y}(z))}.
    \label{eq: GFbackoff}
\end{equation}

The generating function of the channel occupancy of a collision involving AC$_k$ stations is:
\begin{equation}
    \widehat{Y}(z) = (1 - c_k) z^{\sigma / \delta} + \sum_{\ell=1}^I \gamma_{k,\ell} \widehat{G}_\ell(z) + \nu_{k} \widehat{H}(z),
\end{equation}
where \( \gamma_{k,\ell} \) is the probability that AC\(_k\) is blocked by AC\(_\ell\):
\[
\gamma_{k,\ell} = \sum_{j=\max\{Z_k^0,Z_\ell^0\}}^J \frac{\pi(j)}{\sum_{i=Z_k^0}^J \pi(i)} \gamma_{k,\ell}(j),
\]
with $\gamma_{k,\ell}(j)$ defined as:
\[
\gamma_{k,\ell}(j) =
\begin{cases}
    r_k^{n_k-1} n_\ell p_\ell r_\ell^{n_\ell -1} \prod\limits_{\substack{i=1 \\ i \neq k, i \neq \ell}}^ {Z_j} r_i^{n_i}, & \ell \neq k, \\
    (n_k - 1)p_k r_k^{n_k - 2} \prod\limits_{\substack{i=1 \\ i \neq k}}^ {Z_j} r_i^{n_i}, & \ell = k,
\end{cases}
\]
and $\nu_k = c_k - \sum_{\ell=1}^I \gamma_{k,\ell}$ accounts for other collisions. Here, $\pi_j$ is the stationary probability of zone $j$.

The remaining generating functions are given as:
\begin{equation}
    \begin{aligned}
        \widehat{T}(z) &= z^{T_{\text{DATA},k} / \delta}, \\
        \widehat{H}(z) &= \widehat{C}(z) = \widehat{\epsilon}(z)z^{T_\text{C}/\delta}, \\
        \widehat{G}_\ell(z) &= \widehat{\epsilon}(z) z^{(T_{\text{S},\ell} - \text{AIFS}_k - {\text{SIFS}}) / \delta},
    \end{aligned}
\end{equation}
where, \( \widehat{T}(z) \) captures the delay for a successful transmission, $\widehat{C}(z)$ and $\widehat{H}(z)$ capture collision delays, and the function \( \widehat{G}_\ell(z) \) captures the delay experienced by AC\(_k\) due to blocking from AC\(_\ell \).

Based on the total delay generating function in Eq. \ref{eq: total_delay}, the CCDF of the total delay is obtained through numerical inversion of generating function, using the Fourier-series method proposed in \cite{abateIntroductionNumericalTransform2000}:
\begin{equation}
    \begin{aligned}
\Pr(D_k \geq x)
&= \frac1{2\pi r^x}\int_0^{2\pi}\frac{1-\widehat{D}(re^{\mathrm{i}u})e^{-\mathrm{i}xu}}{1 - r e^{\mathrm{i} u}}du \\
&\approx \frac{1}{2N r^x} \sum_{n=-N}^{N-1}
\frac{1-\widehat{D}\left(r e^{-\mathrm{i}\pi n/N}\right)e^{\mathrm{i}\pi n/l}}{1 - r e^{-\mathrm{i}\pi n/N}},
\end{aligned}
\end{equation}
where \( r\in(0,1) \) is a scaling factor,  \( x \) is the integer delay threshold in units of $\delta$, and \( l \) determines the summation range via \( N = xl \). This approximation is accurate for large \( x \), where the tail distribution exhibits exponential decay. As mentioned in \cite{abateIntroductionNumericalTransform2000}, setting \( l = 1 \) and \( r = 10^{-8/v} \) yields numerical errors smaller than \( 10^{-16} \).

To evaluate delay performance, we define the delay reliability index $\theta_k$ for AC$_k$:
\begin{equation}
    \theta_k = -\log \{ \Pr(D_k \geq D_{max,k}) \},
    \label{index}
\end{equation}
where $D_{\max,k}$ is the delay threshold for AC$_k$. A higher $\theta_k$ indicates a lower probability of violating the delay constraint, reflecting greater reliability for QoS-sensitive applications.

\section{Multi-Link EDCA Optimization}
In this section, we present a genetic algorithm-based approach to optimize multi-link EDCA performance by dynamically adjusting contention parameters, aiming to balance delay and packet loss across ACs under varying traffic conditions.

\begin{figure}[t]
    \centering
    \subfloat[]{\includegraphics[width=3.4in]{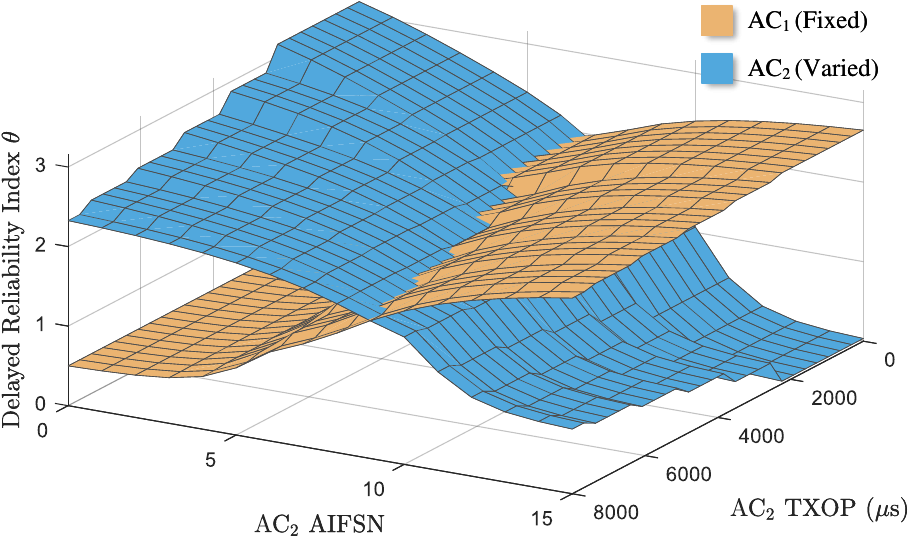}%
    \label{fig_sens_a_index}}
    \hfil
    \subfloat[]{\includegraphics[width=3.4in]{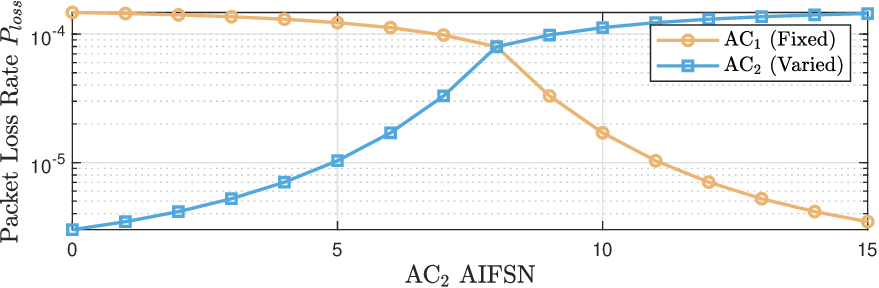}%
    \label{fig_sens_b_loss}}
    \caption{The parameter sensitivity of EDCA: (a) impact to delay reliability index, (b) impact of packet loss probability.}
    \label{fig_sensitivity}
\end{figure}

\subsection{Parameters Sensitivity}
To guide the optimization, we first analyze the sensitivity of QoS metrics to key parameters. In EDCA, parameter selection not only affects performance directly but also influences the interaction between mechanisms. This section analyzes the impact of two newly introduced parameters in EDCA: AIFS and TXOP. And we represent the AIFS by AIFSN (AIFS Number) $= (\text{AIFS}- \text{SIFS})/\sigma$. 

We evaluate a scenario with two ACs: AC$_1$ configured as AIFSN$_1$ = 8 and TXOP$_1$ = 4080~$\mu$s while AC$_2$ has AIFSN$_2$ varying from 2 to 15 and TXOP$_2$ ranging from 0 to 8160~$\mu$s. Other parameters are identical for both AC$_1$ and AC$_2$: CW$_\text{min}$ = 32, CW$_\text{max}$ = 1024, retransmission limit $R$ = 7, number of stations $n$ = 4, packet size $L$ = 1000 bytes, and maximum tolerable delay $D_{max}$ = 100 ms.
Fig.~\ref{fig_sens_a_index} shows the delay reliability index $\theta$ under various AIFSN$_2$ and TXOP$_2$ settings, and Fig.~\ref{fig_sens_b_loss} shows the packet loss probability $P_{loss}$ versus AIFSN$_2$. 

In Fig.~\ref{fig_sens_a_index}, although AC$_1$'s parameters remain fixed, its delay violation probability is affected by AC$_2$’s settings. As AIFSN$_1$ increases, the delay reliability index of AC$_1$ drops, while that of AC$_2$ decrease. Further analysis reveals that TXOP has different effects depending on AIFSN. When AIFSN is small, a large TXOP allows one station to occupy the channel longer, blocking others and increasing delay. In contrast, when AIFSN is large, increasing TXOP mitigates contention and improves delay performance.

Figure~\ref{fig_sens_b_loss} complements this by showing that when AIFSN$_2$ approaches AIFSN$_1$ in Fig.~\ref{fig_sens_a_index}, a balance in delay performance emerges between ACs, a shorter multi-station conflict time will lower the overall delay. However, increased contention leads to fewer blocks but higher collision rates, significantly raising the packet loss probability.

This parameter analysis shows that effective QoS evaluation must consider not only the fulfillment of meeting delay-reliability constraints but also the impact on packet loss to ensure balanced and stable system performance. Optimizing delay alone may increase contention and result in higher packet loss rates.

\subsection{Delay Bound QoS Optimization Problem}
To optimize the QoS performance of EDCA, we formulate a constrained optimization problem. To minimize the total packet loss probability across all ACs while ensuring that each AC’s delay violation probability remains below a specified threshold, the optimization framework treats all ACs equally by default but can be extended to prioritize specific ACs through the incorporation of weighted factors.
\begin{equation}
\begin{aligned}
    \mathbf{P1}:
    \quad
    \arg &\max_{\mathbf{x}_{i},i\in {I_{\text{all}}}} \ 
     \sum_{i = 1}^{I_{\text{all}}}-\log (P_{loss,i})\\
    &\mathrm{s.t.} \ 
     \Pr(D_{i}\geq D_{max,i})<\epsilon_{i},1\leq i\leq I_{\text{all}},
\end{aligned}
\end{equation}
where $\mathbf{x}_{i}$ is the parameter vector of AC$_i$, including CW$_{\min,i}$, CW$_{\max,i}$, AIFSN$_i$, TXOP$_i$, and $R_i$. The constraint ensures that the delay violation probability for each AC does not exceed its threshold $\epsilon_{i}$, addressing the diverse QoS requirements of different ACs.

The retransmission limit $R_i$ is treated as a variable, as it directly impacts both packet loss and delay, as modeled in Eq.~(\ref{eq: Ploss}) and Eq.~(\ref{eq: GFbackoff}). Based on IEEE 802.11 standards, these parameters have pre-defined ranges: CW$_{\min,i}$ and CW$_{\max,i}$ range from 1 to 1023 with CW$_{\min,i}$ $\leq$ CW$_{\max,i}$, AIFSN$_i$ from 2 to 15, TXOP$_i$ from 0 to 8.192 ms with a step size of 32~$\mu$s, and $R_i$ from 4 to 7. These bounds define the feasible parameter space for optimization, balancing delay and loss under varying network conditions.

The optimization problem's large, discrete, and non-convex parameter space renders exhaustive search computationally infeasible. We therefore adopt a genetic algorithm (GA), a powerful heuristic whose population-based search is particularly adept at exploring the vast solution space to avoid local optima. While GA does not guarantee global optimality, its key hyperparameters were empirically tuned for robust performance, and its practical convergence for this problem is demonstrated in our results.

For EDCA parameter tuning, the proposed GA minimizes packet loss while satisfying delay constraints, as outlined in Algorithm~1. It initializes diverse parameter sets, evolves them via selection, crossover, and mutation, and retains elite solutions to converge to a near-optimal configuration. A fitness function prioritizing low packet loss enhances adaptive QoS and robustness in high-load scenarios.

\addtolength{\topmargin}{0.05in}
\begin{algorithm}[t]
    \label{alg:GA}
    \caption{Genetic Algorithm for EDCA Parameter Optimization}
    \begin{algorithmic}[1]
        \State Initialize environment and define parameter ranges for AIFSN, CW$_{\min}$, CW$_{\max}$, TXOP, and R.
        \State Set fixed parameters: ${I_{\text{all}}}$, $n$, $D_{\max}$, $\epsilon$.
        \State Set genetic algorithm parameters: population size $N_{\text{pop}}$, maximum generations $N_{\text{gen}}$, crossover rate $\beta_{\text{cross}}$, elite count $N_{\text{elite}}$, and early stopping threshold $N_{\text{stag}}$.
        \State Define fitness function: compute all $P_{loss,i}$ and return $\sum_{i=1}^{I_{\text{all}}} -\log(P_{loss,i})$.
        \State Define constraints: $\Pr(D \geq D_{\max}) - \epsilon \leq 0$.
        \State Run genetic algorithm:
        \LState Initialize population within parameter bounds.
        \For{each generation}
        \LState  Evaluate fitness for each individual.
        \LState  Apply selection to choose parents.
        \LState  Perform crossover and mutation to generate offspring.
        \LState  Preserve elite individuals.
        \LState  Update population.
        \EndFor
        \LState Decode optimal solution $P_{loss}, \Pr(D \geq D_{\max})$ and output parameters.
    \end{algorithmic}
\end{algorithm}

\section{Results and Discussion}
In this section, we present the simulation results of the proposed QoS optimization method. We compare the performance of three distinct scenarios: the optimized multi-link EDCA parameters, the optimized single-link EDCA parameters, and the single-link EDCA using default parameters.

The parameter settings in the simulation are summarized in Table I unless otherwise specified or varied.
\begin{table}
    \begin{center}
      \caption{Simulation Parameters}
      \begin{tabular}{c|c|l} 
        \hline
        \textbf{Parameter} & \textbf{Value} & \textbf{Definition}\\
        \hline
        ${r}_\text{ctrl}$ & 1 Mbps & Control rate for conventional channel\\
        ${r}_\text{data}$ & 11 Mbps & Data rate for conventional channel\\
        $\sigma$ & 20 $\mu$s & Slot time duration\\
        $\delta$ & 10 $\mu$s & Discrete time slot\\
        $T_\text{SIFS}$ & 10 $\mu$s & SIFS (Short Interframe Space) time\\
        $T_\text{PHY\_H}$ & 192 $\mu$s & Physical layer header time\\
        \hline
        $L_\text{RTS}$ & 160 bits & R-RTS frame length\\
        $L_\text{CTS}$ & 112 bits & R-CTS frame length\\
        $L_\text{ACK}$ & 112 bits & ACK frame length\\
        $L_\text{MAC\_H}$ & 224 bits & MAC layer header length\\
        \hline
        $N_{\text{pop}}$ & 2000 & Population size\\
        $N_{\text{gen}}$ & 500 & Maximum generations\\
        $N_{\text{elite}}$ & 80 & Elite count\\
        $N_{\text{cross}}$ & 0.8 & Crossover rate\\
        $N_{\text{stag}}$ & 50 & Early stopping threshold\\
        \hline
      \end{tabular}
    \end{center}
\end{table}

To evaluate the proposed GA for optimizing multi-link EDCA parameters, we conducted simulations using a BSS with five ACs, each configured with distinct QoS requirements. We set the number of stations for every AC $n = [2, 4, 4, 3, 3]$, packet sizes $L = [50, 210, 256, 800, 2000]$ bytes, maximum tolerable delays $D_{\max} = [50, 60, 100, 300, 300]$ ms, and delay violation probability thresholds $\epsilon_i = [10^{-7}, 10^{-6}, 10^{-4}, 10^{-2}, 0.5]$. These settings reflect a diverse traffic mix, with AC$_1$ and AC$_2$ representing delay-sensitive services, AC$_3$ for streaming, and AC$_4$ and AC$_5$ for best-effort and background traffic.

\begin{figure}[t]
    \centering
    \includegraphics[width=3.4in]{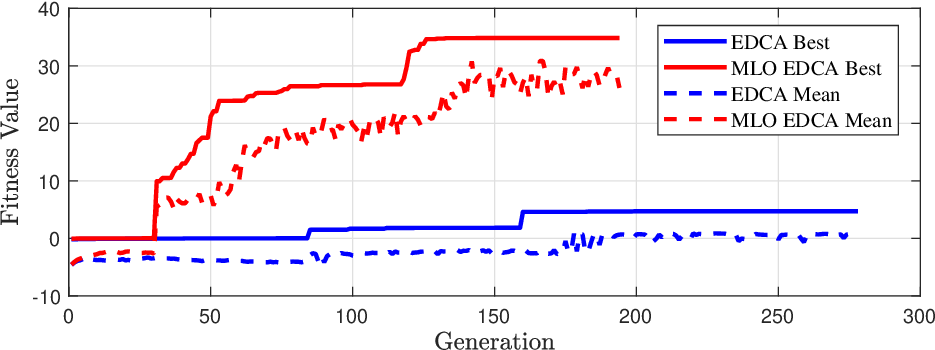}
    \caption{Fitness function convergence for MLO and Single-Link EDCA Optimization.}
    \label{fig:fit-gener}
\end{figure}

Figure~\ref{fig:fit-gener} shows the fitness function convergence over generations for the genetic algorithm applied to both multi-link EDCA optimization (MLO EDCA) and single-link EDCA optimization (EDCA). This figure shows the algorithm’s efficiency in minimizing packet loss under delay constraints. The curves for both best and mean fitness values converge quickly. They stabilize at near-optimal levels.

\begin{figure}[t]
    \centering
    \subfloat[]{\includegraphics[width=1.7in]{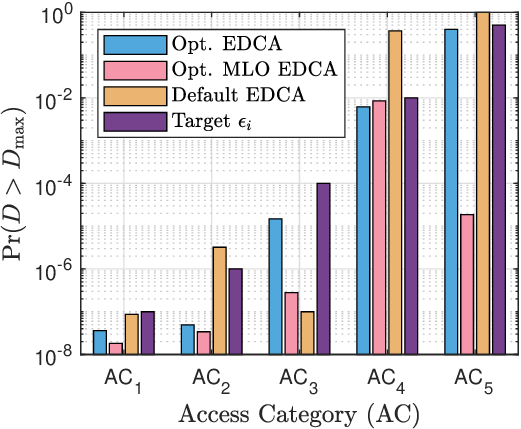}%
    \label{fig_a_a_PrD}}
    \hfil
    \hfil
    \subfloat[]{\includegraphics[width=1.7in]{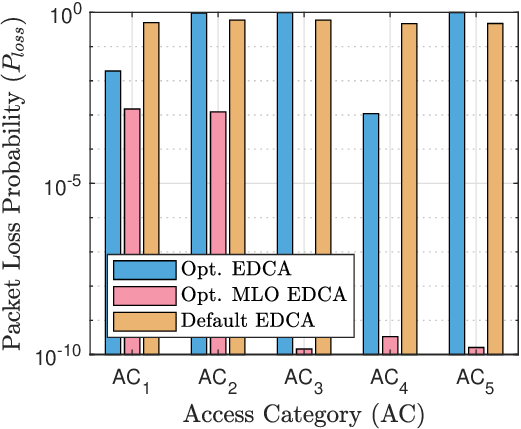}%
    \label{fig_a_b_ploss}}
    \caption{Comparison of QoS performance for Default EDCA, Optimized EDCA, and Optimized MLO EDCA: (a) delay violation probability, (b) packet loss probability.}
    \label{fig_ans}
\end{figure}

Figure~\ref{fig_ans} compares the QoS performance across the five ACs for three distinct configurations: the default single-link EDCA settings, the optimized single-link EDCA parameters found by the GA, and the optimized multi-link EDCA configuration found by the GA. Fig.~\ref{fig_a_a_PrD} illustrates the delay violation probability $\Pr(D \geq D_{\max})$, showing that both optimized configurations significantly reduce violations compared to the default settings. Similarly, Fig.~\ref{fig_a_b_ploss} shows the packet loss probability $P_{\text{loss}}$. The optimized multi-link EDCA configuration significantly outperforms single-link EDCA configurations, achieving the lowest loss rates and ensuring enhanced reliability for time-critical applications.

\begin{figure}[t]
    \centering
    \subfloat[]{\includegraphics[width=1.7in]{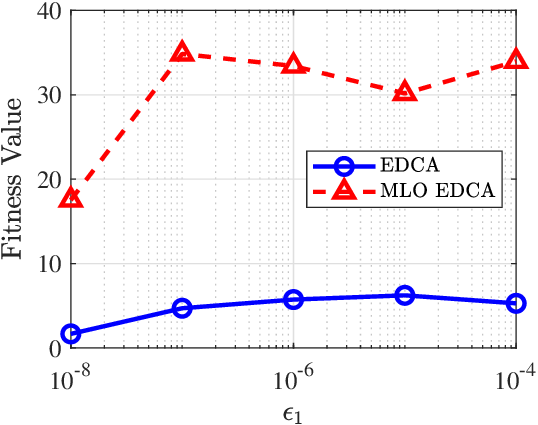}%
    \label{fig_epsl_a_PrD}}
    \hfil
    \hfil
    \subfloat[]{\includegraphics[width=1.7in]{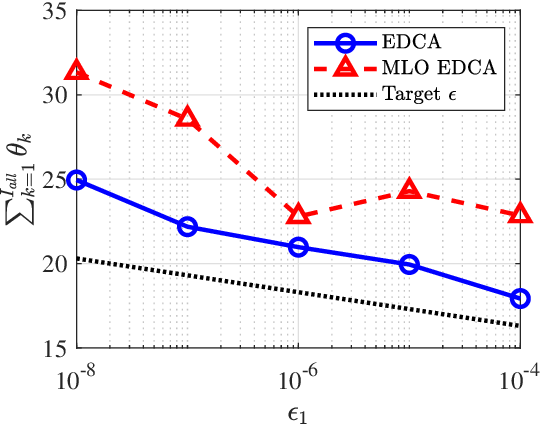}%
    \label{fig_epsl_b_ploss}}
    \caption{Impact of target $\epsilon_1$ on QoS performance of Optimized EDCA and Optimized MLO EDCA: (a) fitness value, (b) sum of reliability index.}
    \label{fig_epsl}
\end{figure}

Figure~\ref{fig_epsl} shows how the target reliability constraint of AC$_1$ ($\epsilon_1$) influences the QoS performance of the optimized EDCA and multi-link EDCA configurations. As shown in Fig.~\ref{fig_epsl}a, the fitness value does not increase monotonically, reflecting a trade-off between fitness and reliability in both Fig.~\ref{fig_epsl}a and Fig.~\ref{fig_epsl}b. Moreover, the fitness value is consistently higher for multi-link EDCA, owing to its ability to leverage multiple links for improved reliability. These findings highlight the complex relationship between performance optimization and reliability constraints in multi-link network configurations.

\section{Conclusion}

In this paper, we addressed the critical challenge of guaranteeing strict delay bounds for QoS in IEEE 802.11be MLO STR networks. We propose an optimization framework utilizing an analytical model and a genetic algorithm. This framework jointly optimizes AC traffic allocation among links and per-link EDCA parameter configurations to minimize overall packet loss while satisfying delay violation probability constraints. Our evaluation demonstrates that this approach effectively identifies viable configurations that satisfy stringent delay constraints while achieving significant performance gains. Future work will explore QoS performance optimization in various MLO working scenarios, including unsaturated burst conditions and hardware-based studies.

\bibliographystyle{IEEEtran}
\bibliography{QoSbbtx.bib}
\end{sloppypar}
\end{document}